\documentclass[aps,pre,twocolumn,groupedaddress]{revtex4-1}
\usepackage{graphicx}
\usepackage{dcolumn}
\usepackage{bm}
\usepackage{amsmath}
\usepackage{amssymb}
\usepackage{color}

\begin{document}

%Title of paper
\title{Membrane shape deformation induced by curvature-inducing proteins consisting of chiral crescent binding and intrinsically disordered domains}

\author{Hiroshi Noguchi}
\email[]{noguchi@issp.u-tokyo.ac.jp}
\affiliation{Institute for Solid State Physics, University of Tokyo, Kashiwa, Chiba 277-8581, Japan}

\date{\today}

\begin{abstract}
Curvature-inducing proteins containing a Bin/Amphiphysin/Rvs domain often have intrinsically disordered domains.
Recent experiments have shown that these disordered chains enhance curvature sensing and generation.
Here, we report on the modification of protein--membrane interactions by disordered chains using meshless membrane simulations.
The protein and bound membrane are modeled together as a chiral crescent protein rod with two excluded-volume chains.
As the chain length increases, the repulsion between them reduces the cluster size of the proteins.
It induces spindle-shaped vesicles and a transition between arc-shaped and circular protein assemblies in a disk-shaped vesicle.
For flat membranes, an intermediate chain length induces  many tubules owing to the repulsion between the protein assemblies, whereas
longer chains promote perpendicular elongation of tubules.
Moreover, protein rods with zero rod curvature and sufficiently long chains stabilize the spherical buds.
For proteins with a negative rod curvature, an intermediate chain length induces a rugged membrane with branched protein assemblies,
whereas longer chains induce the formation of tubules with periodic concave-ring structures.
\end{abstract}

\maketitle

\section{Introduction}

In living cells, the shapes of biomembranes are changed dynamically depending on their functions~\cite{mcma05,suet14}.
For example, spherical vesicles are formed via membrane budding during endocytosis, exocytosis, and intracellular traffic.
Curvature-inducing proteins such as clathrin and coat protein complexes bend membranes into spherical buds~\cite{joha15,bran13,hurl10,mcma11,kaks18}.
In contrast, Bin/Amphiphysin/Rvs (BAR) superfamily proteins have a crescent binding domain to bend the membrane along the domain axis;
 thus, their binding generates cylindrical membrane tubes rather than spherical buds~\cite{mcma05,suet14,joha15,itoh06,masu10,baum11,mim12a,fros08,adam15}.
Many BAR proteins also have intrinsically disordered domains (from $50$ to $400$ residues~\cite{piet13,zeno19}). These domains are unfolded in physiological conditions. They are often removed for {\it in vitro} experiments to investigate 
the interactions between the BAR domains and membranes.
However, recent experiments~\cite{stac12,busc15,zeno18,zeno19,snea19} have shown that such intrinsically disordered domains can significantly influence the membrane shapes.
They changed the length of disordered domains in BAR and other proteins using gene manipulation and revealed that the disordered domains induce curvature sensing
and also the long disordered domains induce the formation of small vesicles.
Thus, it is important to understand the role of disordered domains in membrane remodeling.

A disordered domain behaves like a linear polymer chain in a good solvent~\cite{hofm12},
i.e., its mean radius of gyration scales like $R_{\rm g} \sim {n_{\rm poly}}^{0.6}$,
where $n_{\rm poly}$ is the number of Kuhn segments~\cite{doi86,stro97}.
Interactions between anchored polymer chains and membranes have been intensively studied in theory~\cite{lipo95a,hier96,bick01,mars03,bick06}, 
simulations~\cite{auth03,auth05,wern10,wu13}, and experiments~\cite{tsaf01,tsaf03,akiy03,niko07,koth11}.
Polymer anchoring induces a positive spontaneous curvature of the membrane and increases the bending rigidity. These
relations are analytically derived using Green's function method for a low polymer density (mushroom region) \cite{lipo95a,hier96,bick01} and
scaling methods for a high density (brush region) \cite{lipo95a,hier96,mars03} and are confirmed by Monte Carlo (MC) simulations~\cite{auth03,auth05,wern10}.
In  Ref.~\citenum{zeno18}, 
an MC method similar to that developed in Ref.~\citenum{auth03} was used to analyze experimental data.
Under phase-separation conditions, the polymer anchoring can induce the division of a large domain into smaller domains to gain a larger conformational entropy of chains~\cite{wu13}.
The formation of caveolae-like dimple-shaped membranes by the polymer anchoring was discussed in Ref.~\citenum{evan03a}.
In a poor solvent, the polymer anchoring can induce a negative spontaneous curvature~\cite{wern10,evan03a}.
Experimentally, the formation of spherical buds~\cite{tsaf01,tsaf03,niko07} and membrane tubes~\cite{tsaf03,akiy03,koth11} has been observed.
Thus, the effects of polymer anchoring on membrane deformation are well understood.
However, to the best of our knowledge, the combination with the anisotropic bending of the BAR domains has not been studied yet in theory and simulations.

The aim of this study is to clarify the collective effects of the anisotropic curvature-inducing domains and anchored polymer chains.
Previously, we have simulated anisotropic curvature-inducing proteins  using a meshless membrane method~\cite{nogu22a,nogu14,nogu15b,nogu16,nogu16a,nogu17,nogu17a,nogu19a,nogu22}.
The assembly of crescent rods induces tubulation from a flat membrane~\cite{nogu16,nogu17a}, and the protein chirality enhances this ability~\cite{nogu19a}.
Moreover, vesicles are deformed into disks and polyhedral shapes, since the edges are stabilized by protein assemblies~\cite{nogu14,nogu15b,nogu16}.
In this study, anchored chains were added to the chiral protein rod model~\cite{nogu19a} to investigate how these shape deformations are changed by polymer anchoring to proteins.

The simulation model and method are described in Sec.~\ref{sec:method}.
The results are presented and discussed in Sec.~\ref{sec:results}.
Deformations of the vesicles and flat membranes are described 
in Secs.~\ref{sec:ves} and \ref{sec:flat}, respectively.
Finally, a summary is presented in Sec.~\ref{sec:sum}.

\section{Simulation Model}\label{sec:method}

A fluid membrane is represented by a self-assembled single-layer sheet of $N$ particles.
The position and orientation vectors of the $i$-th particle are ${\bm{r}}_{i}$ and ${\bm{u}}_i$, respectively.
The membrane particles interact with each other via a potential $U=U_{\rm {rep}}+U_{\rm {att}}+U_{\rm {bend}}+U_{\rm {tilt}}$.
The potential $U_{\rm {rep}}$ is an excluded-volume interaction with a diameter $\sigma$ for all pairs of particles.
The effective attractive potential $U_{\rm {att}}$ implicitly represents the interaction involving the solvent.
The details of the meshless membrane model and protein rods are described 
in Ref.~\citenum{shib11} and Refs.~\citenum{nogu14,nogu19a}, respectively.
Except for intrinsically disordered domains, 
the protein and membrane models are identical to those used in Ref.~\citenum{nogu19a}.

The bending and tilt potentials
are given by
 $U_{\rm {bend}}/k_{\rm B}T=(k_{\rm {bend}}/2) \sum_{i<j} ({\bm{u}}_{i} - {\bm{u}}_{j} - C_{\rm {bd}} \hat{\bm{r}}_{i,j} )^2 w_{\rm {cv}}(r_{i,j})$
and $U_{\rm {tilt}}/k_{\rm B}T=(k_{\rm{tilt}}/2) \sum_{i<j} [ ( {\bm{u}}_{i}\cdot \hat{\bm{r}}_{i,j})^2 + ({\bm{u}}_{j}\cdot \hat{\bm{r}}_{i,j})^2  ] w_{\rm {cv}}(r_{i,j})$, respectively,
where ${\bm{r}}_{i,j}={\bm{r}}_{i}-{\bm{r}}_j$, $r_{i,j}=|{\bm{r}}_{i,j}|$,
 $\hat{\bm{r}}_{i,j}={\bm{r}}_{i,j}/r_{i,j}$, $w_{\rm {cv}}(r_{i,j})$ is a weight function, and $k_{\rm B}T$ is the thermal energy.
The spontaneous curvature $C_0$ of the membrane is 
given by $C_0\sigma= C_{\rm {bd}}/2$. \cite{shib11}
In this study, $C_0=0$ and $k_{\rm {bend}}=k_{\rm{tilt}}=10$ are used except for the membrane particles belonging to the protein rods.

Using this parameter set,
the bare membrane has a bending rigidity $\kappa/k_{\rm B}T=15 \pm 1$,
area of the tensionless membrane per particle $a_0/\sigma^2=1.2778\pm 0.0002$,
area compression modulus $K_A\sigma^2/k_{\rm B}T=83.1 \pm 0.4$,
 edge line tension $\Gamma\sigma/k_{\rm B}T= 5.73 \pm 0.04$~\cite{nogu14},
and the Gaussian modulus $\bar{\kappa}/\kappa=-0.9\pm 0.1$~\cite{nogu19}.
These are typical values for lipid membranes.

\begin{figure}[t]
\includegraphics[]{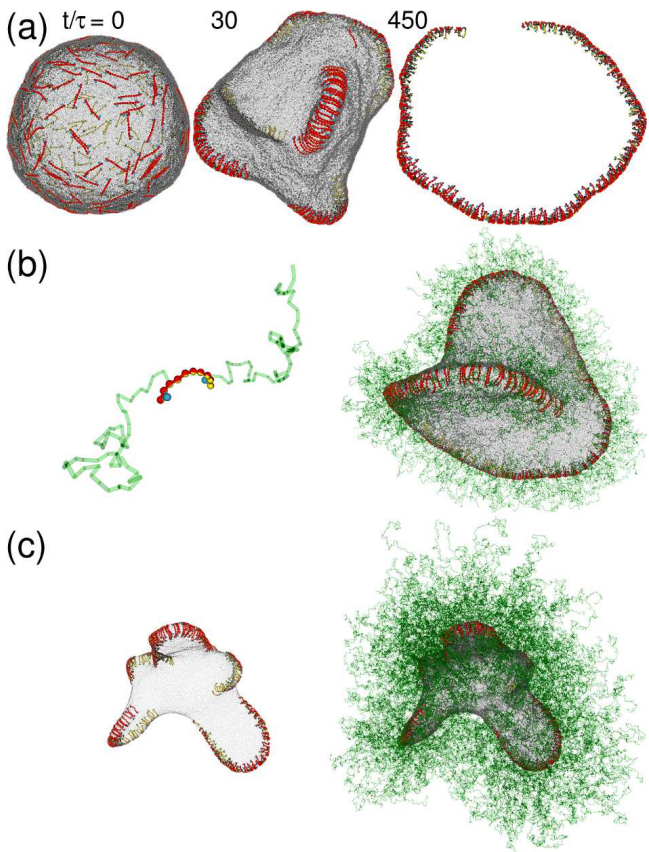}
\caption{
Snapshots of vesicles at $C_{\rm rod}r_{\rm rod}=3$  and $\phi=0.1$.
A protein is displayed as a crescent rod (red and yellow)
connected with two particles (cyan) on the laterally opposite sides
and with two excluded-volume chains (transparent green) consisting of $n_{\rm poly}$ particles, as shown in the left panel in (b).
The orientation vector ${\bf u}_i$ lies along the direction from the yellow to red hemispheres.
The cylinder width and transparency of anchored chains are adjusted for each snapshot for clarity.
The membrane particles are displayed as transparent gray spheres.
Spherical vesicles at $C_{\rm rod}=0$ are used as initial states.
(a) Time evolution for the protein rods with no anchored chains ($n_{\rm poly}=0$).
From left to right, $t/\tau=0$, $30$, and $450$.
The membrane particles are not displayed in the right snapshot for clarity.
(b) A  vesicle shape at $t/\tau=800$ for $n_{\rm poly}=50$.
(c) A vesicle shape at  $t/\tau=800$ for $n_{\rm poly}=200$.
In the left snapshot,
the anchored chains are not displayed and the membrane particles are displayed as small spheres for clarity.
}
\label{fig:snapves}
\end{figure}

A BAR domain and the membrane underneath it are modeled together as a rod,
which is a linear chain of $n_{\rm {sg}}$ membrane particles~\cite{nogu14} (see the left snapshot in Fig.~\ref{fig:snapves}(b)).
We use $n_{\rm {sg}}=10$, i.e., the rod length $r_{\rm rod}=10\sigma$.
The protein rods have spontaneous curvatures $C_{\rm {rod}}$ along the rod axis
and no spontaneous (side) curvatures perpendicular to the rod axis.
The protein-bound membrane is more rigid than the bare membrane as $k_{\rm {bend}}=k_{\rm{tilt}}=80$.
Since two identical binding domains of BAR proteins form a dimer that has a two-fold rotational symmetry and chiral, we model it as follows.
Two particles are connected to the last two particles next to the chain ends by a bond potential
and located laterally on opposite sides to represent the protein chirality~\cite{nogu19a}
(see cyan particles in Fig.~\ref{fig:snapves}(b)). 
Two linear chains consisting of $n_{\rm poly}$ particles are anchored to the third rod particles from both ends
(see green particles in Fig.~\ref{fig:snapves}(b)). 
These protein particles have excluded-volume interactions with  diameter $\sigma$ as the membrane particles.
Neighboring segments in the chains are connected by a harmonic potential $U_{\rm bond}= (k_{\rm poly}/2)(r_{ij}- r_{\rm p0})^2$.
In this study, $k_{\rm poly}=10k_{\rm B}T$ and $r_{\rm p0}=1.2\sigma$ are used.
These excluded-volume chains are set on one side of the membrane (outside of a vesicle or upper side of a flat membrane).

We primarily use a rod curvature of $C_{\rm rod}r_{\rm rod}=3$ that corresponds to that of BAR-PH~\cite{zhu07}.
The length of the excluded-volume chains are varied as $n_{\rm poly}=0$, $10$, $25$, $50$, $100$, and $200$:
the mean end-to-end distance of a free chain is $r_{\rm end} \simeq r_{\rm p0}{n_{\rm poly}}^{0.6}= 0$, 
$4.8\sigma$, $8.3\sigma$, $13\sigma$, $19\sigma$, and $29\sigma$, respectively,
where the exponent $0.6$ is for the excluded-volume chains in a good solvent~\cite{doi86,stro97}.
Since the Kuhn length of unfolded proteins is $\simeq 0.8$\,nm (approximately two residues)~\cite{hofm12},
$n_{\rm poly}=100$ corresponds to a disordered domain of $200$ residues.
The excluded-volume chains are attached to all protein rods unless  otherwise specified.
For tubulation from a flat membrane, the case of the excluded-volume chains attached only to 10\% of protein rods is also examined.
The total number of membrane particles $N=25600$ is used for both vesicles and flat membranes.
The radius of the spherical vesicle is $50\sigma$.
The protein density varies from $\phi= n_{\rm {sg}}N_{\rm {rod}}/N=0.0375$ to $0.1$ for the vesicles
and $\phi=0.05$ and $0.2$ for flat membranes, respectively.
Zero surface tension is used for the flat membranes.
To generate an initial conformation,
straight chains are anchored first and then the chain conformation is relaxed with or without fixing the membrane position.
Molecular dynamics with a Langevin thermostat is employed~\cite{shib11,nogu11}.
For the time unit, $\tau= r_{\rm {rod}}^2/D$ is used,
where $D$ is the diffusion coefficient of the membrane particles in the tensionless membrane~\cite{nogu16a}.
Error bars are estimated from three and ten independent runs for vesicles and tubulations, respectively.

\begin{figure}[t]
\includegraphics[]{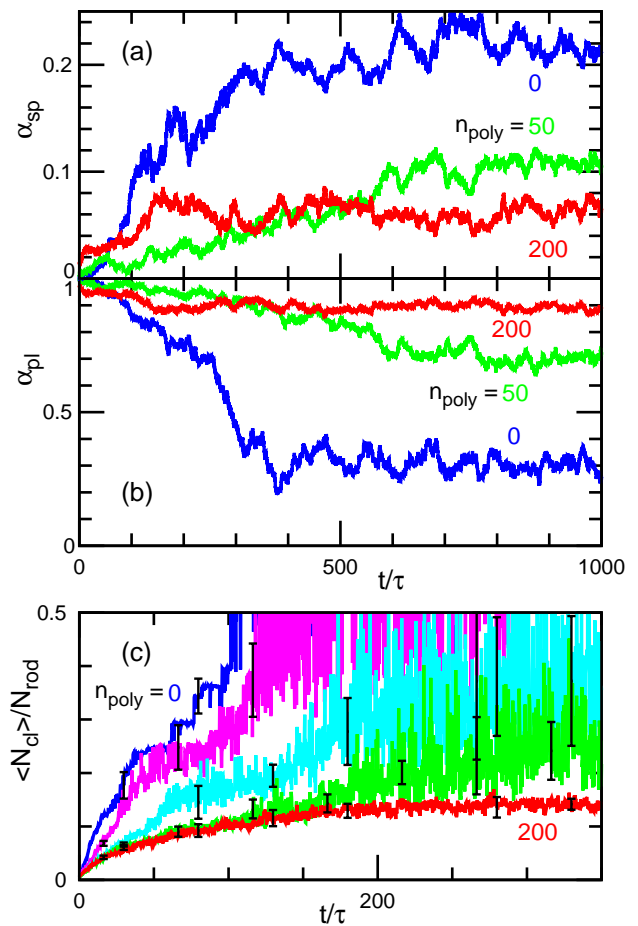}
\caption{
Time development of (a) the asphericity $\alpha_{\rm sp}$, (b) aplanarity $\alpha_{\rm pl}$, and (c) mean cluster size $\langle N_{\rm cl} \rangle$   at $C_{\rm rod}r_{\rm rod}=3$  and $\phi=0.1$.
The blue, green, and red curves in (a),(b) correspond to those in Figs.~\ref{fig:snapves}(a), (b), and (c), respectively.
(c) The cluster size is averaged for three independent runs, 
and the error bars are displayed at several data points.
From top to bottom, $n_{\rm poly}=0$, $10$, $25$, $50$, and $200$.
}
\label{fig:sphrg}
\end{figure}

A protein is considered to belong to a cluster
when the distance between the centers of mass of the rod region
and one of the rods in the cluster is less than $r_{\rm {rod}}/2$. 
The mean size of the clusters is given by
$N_{\rm {cl}}= (\sum_{i_{\rm {cl}}=1}^{N_{\rm {rod}}} i_{\rm {cl}}^2 n^{{\rm {cl}}}_i)/N_{\rm {rod}}$,
where $n^{{\rm {cl}}}_i$  is the number of clusters of size $i_{\rm {cl}}$.

To quantify vesicle shapes, we calculate two shape parameters based on
the gyration tensor $a_{\alpha\beta}= (1/N)\sum_i (\alpha_{i}-\alpha_{\rm G})(\beta_{i}-\beta_{\rm G})$
of membrane particles including the bound region of the proteins (i.e., excluding anchored chains), where $\alpha,\beta\in \{x,y,z\}$:
\begin{eqnarray}
\alpha_{\rm {sp}} &=& \frac{({\lambda_1}-{\lambda_2})^2 + 
  ({\lambda_2}-{\lambda_3})^2+({\lambda_3}-{\lambda_1})^2}{2(\lambda_1+\lambda_2+\lambda_3)^2}, \\ 
\alpha_{\rm {pl}} &=&  \frac{9\lambda_1\lambda_2\lambda_3} {(\lambda_1+\lambda_2+\lambda_3)
    (\lambda_1\lambda_2+\lambda_2\lambda_3+\lambda_3\lambda_1)},
\end{eqnarray}
where ${\lambda_1} \leq {\lambda_2} \leq {\lambda_3}$ are 
three eigenvalues of the gyration tensor.
Spherical, disk, and rod shapes are well distinguished by $\alpha_{\rm {sp}}$ and $\alpha_{\rm {pl}}$. 
The asphericity, $\alpha_{\rm {sp}}$, indicates the shape deformation from a spherical 
shape to a thin-rod shape:~\cite{rudn86}  $\alpha_{\rm {sp}}=0$, $0.25$, and $1$ 
for a sphere (${\lambda_1}={\lambda_2}={\lambda_3}$), thin disk (${\lambda_1}=0, {\lambda_2}={\lambda_3}$), 
and thin rod (${\lambda_1}={\lambda_2}=0$), respectively.
A discocyte (red blood cell) shape takes $\alpha_{\rm {sp}}\simeq 0.2$.~\cite{nogu05}
The aplanarity, $\alpha_{\rm {pl}}$, indicates the deviation from a plane:~\cite{nogu06}
 $\alpha_{\rm {pl}}=0$  and $1$ for a flat membrane (${\lambda_1}=0$) and sphere, respectively.
In addition to vesicles~\cite{nogu05,nogu15a,tame20} and polymer chains~\cite{nogu98},
these quantities were used to characterize the cup-shaped membrane closing into a vesicle~\cite{nogu19} and
the arrangements of pores of the vesicle~\cite{nogu16b}.

\begin{figure}[t]
\includegraphics[]{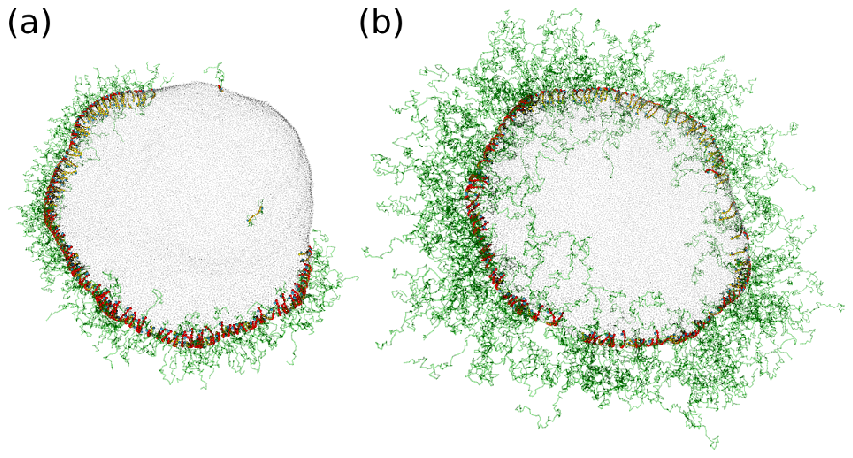}
\includegraphics[]{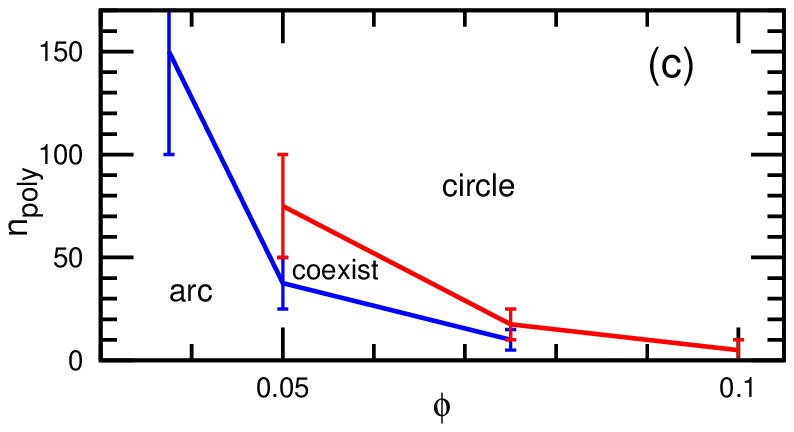}
\caption{
Transition between arc-shaped and circular assemblies of protein rods at  $C_{\rm rod}r_{\rm rod}=3$.
(a),(b) Snapshots of (a) the arc-shaped and (b) circular assemblies for $n_{\rm poly}=25$ and $100$  at $\phi=0.05$, respectively.
The membrane particles are displayed as small transparent spheres for clarity.
(c) Phase diagram. The circular (arc-shaped) assembly exists in the region above the blue line (under the red line), 
when the circular (arc-shaped) assembly is used as an initial conformation. These two states coexist in the region between two lines.
}
\label{fig:pd}
\end{figure}

\begin{figure}[t]
\includegraphics[]{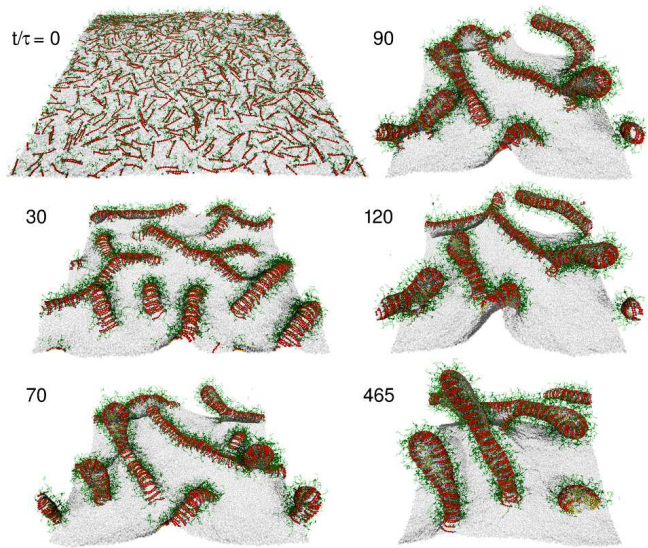}
\caption{
Sequential snapshots of tubulation from a flat membrane at $C_{\rm rod}r_{\rm rod}=3$, $\phi=0.2$, and $n_{\rm poly}=10$.
From top left to bottom right, $t/\tau=0$, $30$, $70$, $90$, $120$, and $465$.
}
\label{fig:snapn512n10}
\end{figure}

\section{Results and Discussion}\label{sec:results}

\subsection{Vesicle Deformation}\label{sec:ves}

First, we consider the deformation of vesicles by the proteins (Figs.~\ref{fig:snapves}--\ref{fig:pd}).
At $C_0=0$, the proteins are randomly distributed on the vesicle either with or without anchored chains (see the left snapshot in Fig.~\ref{fig:snapves}(a)).
As the curvature is changed to $C_0r_{\rm rod}=3$ at $t=0$, the proteins self-assemble on the vesicle from this random state (Figs.~\ref{fig:snapves} and \ref{fig:sphrg}).
Before discussing the effects of the anchored chains, we describe the vesicle deformation in their absence ($n_{\rm poly}=0$).
A parallel pair of proteins have an attraction in the direction perpendicular to the protein axis via membrane-mediated interactions~\cite{nogu17} so that the proteins form a side-by-side assembly along one direction.
At a sufficiently low protein density $\phi$, the protein assembly exhibits an arc shape, leading to a disk-shaped vesicle~\cite{nogu14} (see Figs.~\ref{fig:snapves}(a)--(c) and Movie S1). At a higher density, the arc is closed to a circular shape, and with increasing density the assembly elongates to an elliptic shape~\cite{nogu15b,nogu16}. At an even higher density, polyhedral vesicles are formed~\cite{nogu15b}.
Note that the chirality gives negligible effects on these straight assemblies~\cite{nogu19a}.

Anchored chains change the vesicle shape.
As anchored chains become longer at $\phi=0.1$, vesicles exhibit a twisted-arc shape at $n_{\rm poly}= 50$ (Fig.~\ref{fig:snapves}(b)) and  multiple arc-shaped protrusions at $n_{\rm poly}\gtrsim 100$ (Fig.~\ref{fig:snapves}(c) and Movie S2).
Although this twisted-arc shape can be temporally formed by proteins without chains at a higher density~\cite{nogu16},
it quickly transforms into a disk shape. Conversely, the twisted shape is maintained for a long period (at least until $t/\tau=1000$) at $n_{\rm poly}= 50$ owing to the chain interactions.
The anchored chains generate repulsion between neighboring proteins as well as between protein assemblies.
Moreover, they induce an effective spontaneous curvature, such that the protein assembly is more curved along the arc.
These interactions suppress the fusion of protein assemblies and stabilize the shape, as shown in Fig.~\ref{fig:snapves}(c). 
To investigate the assembly dynamics in more details, the time development of the vesicle shapes and protein cluster sizes is plotted in Fig.~\ref{fig:sphrg}.
The disk-shaped vesicle with no anchored chains ($n_{\rm poly}=0$) has the asphericity $\alpha_{\rm sp} \simeq 0.2$ and aplanarity $\alpha_{\rm pl} \simeq 0.3$. 
The chain anchoring makes vesicles more spherical shaped in the ellipsoidal approximation
(Figs.~\ref{fig:sphrg}(a) and (b)).
The initial stage of the protein assembly becomes slower as the chain length increases from $n_{\rm poly}=0$ to $50$,
and it is almost constant at $n_{\rm poly} \geq 50$ (see Fig.~\ref{fig:sphrg}(c)).
This threshold length $n_{\rm poly}=50$, the average distance ($\simeq 10\sigma$) between randomly distributed protein rods is close to the end-to-end distance of free chains.
Hence, the initial assembly is slowed by the repulsive interaction between the anchored chains,
but this effect is saturated at a chain length equal to the protein distance.

To clarify the interaction between neighboring protein rods in the assembly,
the closing transition of the arc-shaped assembly is investigated (Fig.~\ref{fig:pd}).
When long chains are anchored to an arc-shaped assembly,
the assembly transforms into a circular shape, as shown in Fig.~\ref{fig:pd}(b).
Conversely, when the chain length is reduced,
the circular assembly opens into an arc shape, as shown in Fig.~\ref{fig:pd}(a).
The thresholds of these changes are different, as shown in Fig.~\ref{fig:pd}(c),
and the two states coexist at intermediate chain lengths.
Therefore, this is a discrete transition.
The mean distance between neighboring protein rods increases with increasing chain length.
Even for low density $\phi\simeq 0.04$, where less than a half of the circumference is occupied by the protein rods, 
a circular assembly is formed at long chains (Fig.~\ref{fig:pd}(c)).

\begin{figure}[t]
\includegraphics[]{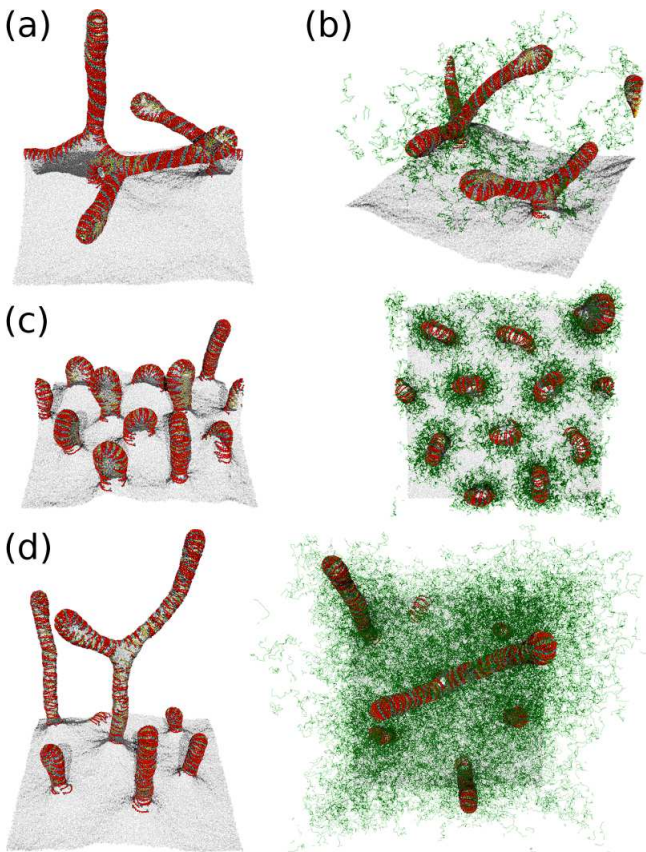}
\caption{
Snapshots of tubulation from a flat membrane at $C_{\rm rod}r_{\rm rod}=3$ and $\phi=0.2$
(a) $n_{\rm poly}=0$. (b) 10\% of the protein rods have anchored chains of $n_{\rm poly}=100$.
(c)  $n_{\rm poly}=25$. (d)  $n_{\rm poly}=100$.
The anchored chains are not displayed in the left snapshots in (c) and (d) for clarity.
The right snapshots in (c) and (d) are from the top view.
}
\label{fig:snapn512s}
\end{figure}

\begin{figure}[t]
\includegraphics[]{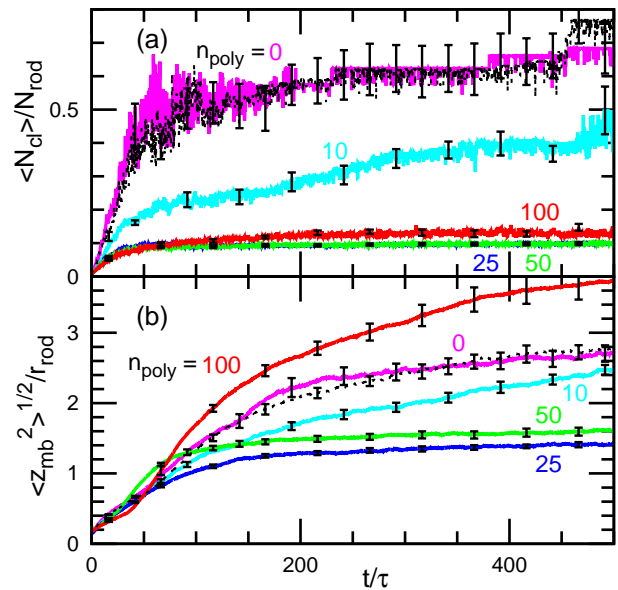}
\caption{
Time development of (a) mean cluster size $\langle N_{\rm {cl}} \rangle$ and 
(b) vertical membrane span  $\langle {z_{\rm {mb}}}^2 \rangle^{1/2}$ at $C_{\rm rod}r_{\rm rod}=3$ and $\phi=0.2$.
The solid lines represent the data of the chains anchoring to all protein rods for $n_{\rm poly}=0$, $10$, $25$, $50$, and $100$.
The dashed lines represent the data of the chains anchoring to 10\% of protein rods for $n_{\rm poly}=100$.
}
\label{fig:flatn512}
\end{figure}

\begin{figure}[t]
\includegraphics[]{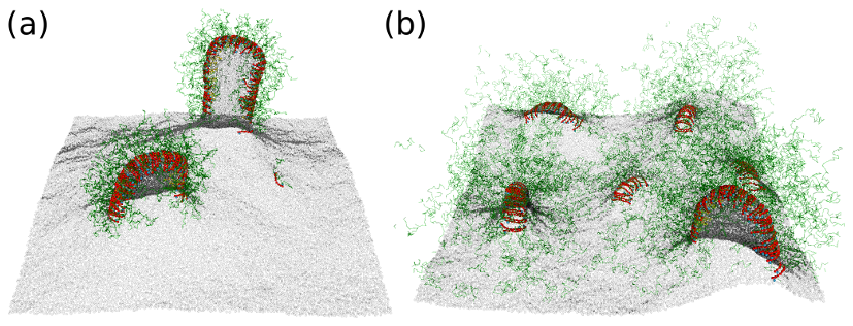}
\includegraphics[]{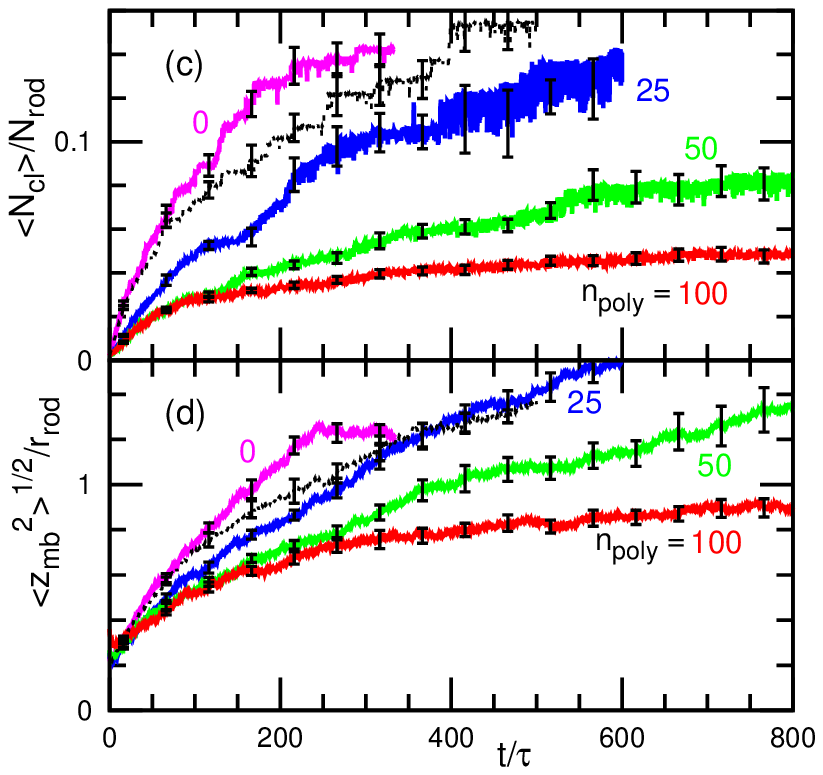}
\caption{
Tubulation from a flat membrane at $C_{\rm rod}r_{\rm rod}=3$ and low density $\phi=0.05$.
(a),(b) Snapshots for (a) $n_{\rm poly}=25$ and (b) $n_{\rm poly}=100$ at $t/\tau=600$ and $800$, respectively.
(c),(d) Time development of (c) mean cluster size $\langle N_{\rm {cl}} \rangle$ and 
(d) vertical membrane span  $\langle {z_{\rm {mb}}}^2 \rangle^{1/2}$.
The solid lines represent the data of the chains anchoring to all protein rods for $n_{\rm poly}=0$, $25$, $50$, and $100$.
The dashed lines represent the data of the chains anchoring to 10\% of protein rods for $n_{\rm poly}=100$.
}
\label{fig:flatn128}
\end{figure}

\subsection{Deformation of Flat Membrane}\label{sec:flat}

Next, we consider tubulation from a flat tensionless membrane (Figs.~\ref{fig:snapn512n10}--\ref{fig:flatn128}).
At high density ($\phi=0.2$), tubulation occurs for all values of $n_{\rm poly}$ at $C_{\rm rod}r_{\rm rod}=3$.
For short chains ($n_{\rm poly} \leq 10$),  tubules are formed via bending of the protein assembly and fuse into a larger tubule 
(see Fig.~\ref{fig:snapn512n10} and Movie S3 for $n_{\rm poly} =10$ and Fig.~\ref{fig:snapn512s}(a) for $n_{\rm poly}=0$).
The protein chirality promotes tubulation via helical assembly on tubules~\cite{nogu19a}.
With increasing chain length $n_{\rm poly}$, the mean cluster size  $\langle N_{\rm {cl}} \rangle$ and membrane vertical span $\langle {z_{\rm {mb}}}^2 \rangle^{1/2}$ decrease until $n_{\rm poly}=25$ (Fig.~\ref{fig:flatn512}).
The vertical span of the membrane is calculated from 
the membrane height variance as 
${z_{\rm mb}}^2=\sum_{i}^{N} (z_i-z_{\rm G})^2/N$,
where $z_{\rm G}$ is the center of mass in the vertical ($z$) direction.
Interestingly, the vertical span increases from  $n_{\rm poly}=25$ to  $n_{\rm poly}=100$.
At  $n_{\rm poly}=25$, the short tubules are separated by the repulsive interaction between anchored chains,
where the distance between tubules is greater than the end-to-end distance $r_{\rm end}$ (see the right snapshot in Fig.~\ref{fig:snapn512s}(c)).
A similar stabilization of small domains was previously obtained in a phase-separated membrane, in which polymer chains were anchored to one of the phases~\cite{wu13}.
At  $n_{\rm poly}=100$, $r_{\rm end}$ is greater than the tubule distance,
so that the chains push the tubules in the upper direction and also lead to narrower shapes to gain more conformational entropy.
Hence, vertically elongated tubules are obtained as shown in  Fig.~\ref{fig:snapn512s}(d) and Movie S4.

For the low density ($\phi=0.05$), tubulation is suppressed with increasing chain length from  $n_{\rm poly}=0$ to  $n_{\rm poly}=100$ (see Fig.~\ref{fig:flatn128}). 
Since fewer protein assemblies are more separated,
the dependency for $n_{\rm poly}\leq 25$ at $\phi=0.2$ is seen even at $n_{\rm poly}=100$.

Until this point, the chains are anchored to all proteins.
When the chains of $n_{\rm poly}=100$ are anchored to 10\% of the proteins,
their effects are negligibly small.
For $\phi=0.2$, the difference from the unanchored proteins is not recognized (compare the  magenta and black dashed lines in Fig.~\ref{fig:flatn512}, and also compare the snapshots in Figs.~\ref{fig:snapn512s}(a) and (b)).
For $\phi=0.05$, there are small differences after tubules are formed (compare the  magenta and black dashed lines in Fig.~\ref{fig:flatn128}).
Thus, the interactions between anchored chains are essential for the tubulation, 
whereas the effects of isolated chains are marginal.

\begin{figure}[t]
\includegraphics[]{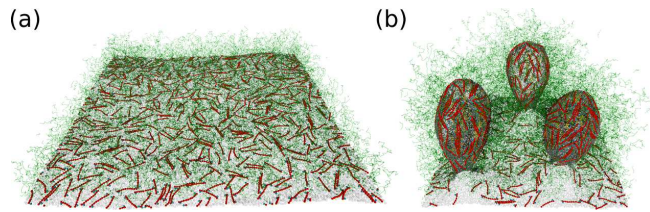}
\caption{
Snapshots of flat membranes at $C_{\rm rod}=0$, $\phi=0.2$, and  $n_{\rm poly}=50$.
(a) Randomly distributed proteins in a flat membrane.
(b) Formation of spherical buds. 
}
\label{fig:snapc0}
\end{figure}

Next, we consider the formation of spherical buds.
At $C_{\rm rod}=0$, the flat membrane is stable on our simulation time scale ($\sim 1000\tau$, see Fig.~\ref{fig:snapc0}(a)).
However, when a tubulated membrane is used as an initial state,
one to three spherical buds are formed at  $n_{\rm poly}=50$ (see Fig.~\ref{fig:snapc0}(b) and Movie S5).
Note that the bud shapes deviate form a perfect sphere.
The membrane is elongated in the vertical direction to gain the conformational entropy of polymer chains,
 and the protein rods are orientated in the vertical direction. 
For $n_{\rm poly}=100$, this elongation is enhanced.
When the chain length is reduced, one large bud remains  at $n_{\rm poly}=25$, but all buds disappear at  $n_{\rm poly}=10$.
Thus, the chain anchoring can induce buds, although a high free-energy barrier exists for their formation.

\begin{figure}[t]
\includegraphics[]{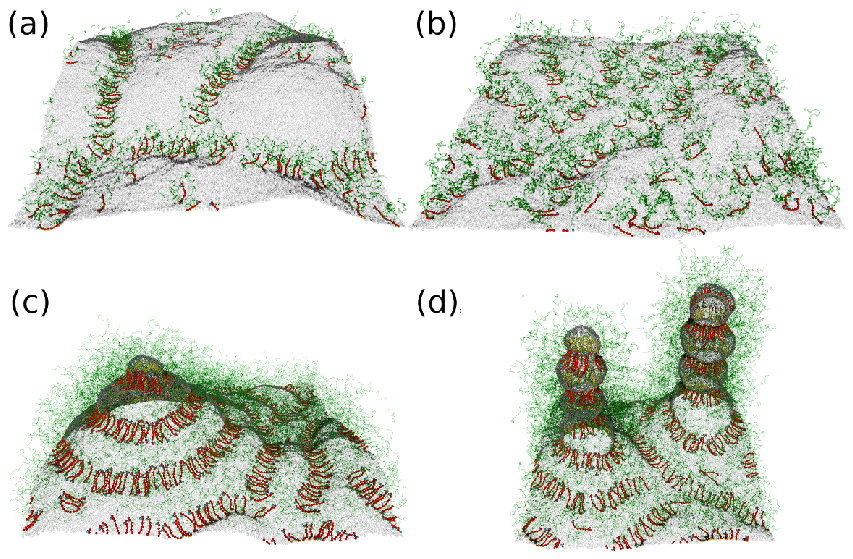}
\includegraphics[]{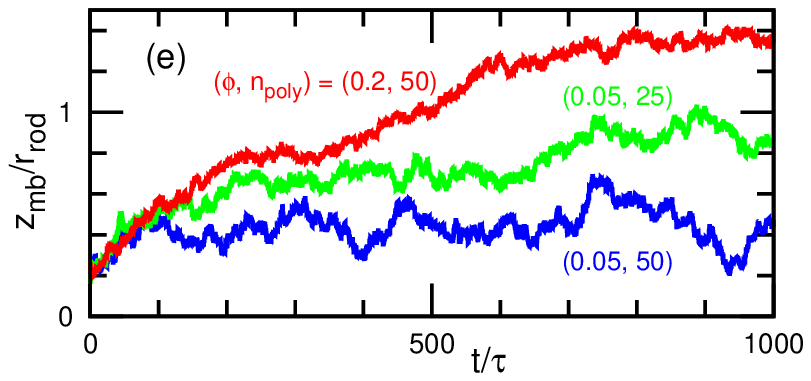}
\caption{
Deformation of flat membranes at $C_{\rm rod}r_{\rm rod}=-3$.
Excluded-volume chains are anchored to the concave side of protein rods.
(a)--(d) Snapshots of membranes at (a) $(\phi, n_{\rm poly})=(0.05,25)$,
(b) $(\phi, n_{\rm poly})=(0.05,50)$, and (c),(d) $(\phi, n_{\rm poly})=(0.2,50)$.
Randomly mixed states at $C_{\rm rod}=0$ are used as initial states for (a)--(c),
while a tubulated state at $C_{\rm rod}r_{\rm rod}=3$ is used for (d).
(e) Time development of the vertical span of the membrane. 
Top to bottom: $(\phi, n_{\rm poly})=(0.2,50)$, $(0.05,25)$, and $(0.05,50)$.
The snapshots in (a)--(c) correspond to the data at $t/\tau=800$.
}
\label{fig:cn6}
\end{figure}

Finally, we consider the case in which protein rods and anchored chains bend the membrane in opposite directions,
i.e., the chains are anchored to the concave side of the rods ($C_{\rm rod}<0$).
When these opposite bending effects are compatible,
proteins form branched assemblies on the membrane as shown in Fig.~\ref{fig:cn6}(a).
Thus, the membrane is rugged by these bumped assemblies.
This behavior agrees well with the experimental observation of vesicles induced by a chimeric protein of I-BAR and disordered domains~\cite{snea19}.
They showed that the disordered domain suppressed inward membrane invagination, and vesicles became a rugged shape.
For longer chains ($n_{\rm poly}=50$), the anchored chains have stronger bending effects;
proteins are distributed more randomly at the low density ($\phi=0.05$)
and form shish-kebab-shaped tubules, in which concave ring-shaped rod assemblies are formed periodically, at the high density ($\phi=0.2$), as shown in Figs.~\ref{fig:cn6}(b)--(d).
The membrane vertical span $z_{\rm mb}$ increases accordingly (Fig.~\ref{fig:cn6}(e)).
The formation of this tubule is very slow and it continues to grow at $t/\tau\simeq 1000$ (see Figs.~\ref{fig:cn6}(c),(e) and Movie S6).
To examine longer tubular structures,
 spherical buds are used as an initial state and the shish-kebab-shaped tubules are obtained (see Fig.~\ref{fig:cn6}(d)).
Small buds and tubules disappear or merge into larger tubules, as shown in Movie S7.
Although a single ring assembly was previously obtained at the neck of a spherical bud~\cite{nogu17a},
the periodic ring structure was found here for the first time.
Since I-BAR proteins do not contain long disordered domains,
this shish-kebab-shaped tubule is not likely to exist in living cells
but can be experimentally constructed by designing chimeric proteins of I-BAR and longer disordered domains.

\section{Summary}\label{sec:sum}

We have studied the membrane deformation induced by the proteins consisting of a chiral crescent domain and two anchored chains.
We have shown that these two types of domains can bend membranes both cooperatively and uncooperatively.
Anchored chains induce an effective spontaneous curvature of the membrane but also generate repulsion between protein rods as well as between protein assemblies.
The latter can weaken or suppress the membrane deformation by the protein rods, depending on the conditions.
For vesicle deformation, long chains induce multi-spindle vesicles and transition from arc-shaped protein assembly to circular assembly.
For tubulation from a flat membrane, small tubules are formed by repulsion between the crowded chains of tubules.
However, long vertical tubules are formed by very long chains that are longer than the mean distance between tubules. 
Thus, the chains suppress the lateral assembly but promote vertical elongation.
Spherical buds can be formed by straight protein rods with sufficiently long anchored chains.
When the rod and chain domains bend membranes in opposite directions at similar amplitudes,
bumped network assemblies are formed on the membrane.
In contrast, shish-kebab-shaped tubules are formed when the long chains have a stronger bending ability.

Our results suggest that the entropic (crowding) interactions between chains are significant to induce membrane deformation.
The highly packed brush states of the chains induce large repulsion between protein assemblies and membrane bending.
In particular, brushed chains strongly push tubules away from the membrane, when the membrane is completely covered by the chain crowds.
Our findings provide new insights into the mechanism of membrane shape regulation by curvature-inducing proteins with disordered domains
and the design of curvature-inducing proteins. 

\begin{acknowledgments}
This work was supported by JSPS KAKENHI Grant Number JP21K03481.
\end{acknowledgments}

\end{document}